\documentclass{PoS}

\title{Inclusive $\psi$(2S) production at forward rapidity in pp, p-Pb and Pb-Pb collisions with ALICE at the LHC}

\ShortTitle{Inclusive $\psi$(2S) production at forward rapidity in pp, p-Pb and Pb-Pb collisions}

\author{\speaker{Biswarup Paul} %
         \thanks{on behalf of the ALICE Collaboration}\\
       \\ Saha Institute of Nuclear Physics, Kolkata, India\\
        E-mail: \email{biswarup.paul@cern.ch}}

\abstract{The ALICE Collaboration has studied inclusive $\psi$(2S) production in pp, p-Pb and Pb-Pb collisions with the ALICE Muon Spectrometer which covers
the rapidity range 2.5 $<$ $y_{\rm lab}$ $<$ 4. The $\psi$(2S) measurement was performed in the dimuon decay channel. The $\psi$(2S) production cross-section and $\psi$(2S) to J/$\psi$ cross-section ratio in pp collisions will be presented, both integrated and differential in rapidity and in transverse momentum. In p-Pb collisions, $\psi$(2S) results will be compared to the J/$\psi$ ones by means of the production cross-section ratio and the double ratio [$\psi$(2S)/J/$\psi$]$_{{\mathrm {pPb}}}$/[$\psi$(2S)/J/$\psi$]$_{{\mathrm {pp}}}$ studied as a function of the resonance transverse momentum and event activity of the collision. The $\psi$(2S) nuclear modification factor, $R_{{\mathrm {pPb}}}$, will also be presented. Theoretical models based on nuclear shadowing, coherent energy loss or both cannot describe our results. Therefore other mechanisms must be invoked in order to describe the $\psi$(2S) production. Finally, results on $\psi$(2S) production in Pb-Pb collisions will be shown in two transverse momentum ranges as a function of centrality. 
}

\FullConference{ 7th International Conference on Physics and Astrophysics of Quark Gluon Plasma\\
                 1-5 February , 2015 \\
                 Kolkata, India}

\begin{document}

\section{Introduction}
The suppression of quarkonia (bound states of a heavy quark and its anti-quark) in ultra relativistic heavy ion collisions is one of the most prominent probes used to investigate and quantify the properties of the quark gluon plasma (QGP). The in-medium dissociation probability of the different quarkonium states could provide an estimate of the temperature of the system since the dissociation is expected to take place when the medium reaches or exceeds the critical temperature of the phase transition ($T_{\rm c}$), depending on the binding energy of the quarkonium state. For charmonium ($c\overline c$) states, the J/$\psi$ is likely to survive significantly above $T_{\rm c}$ (1.5 - 2 $T_{\rm c}$) whereas $\chi_{\rm c}$ and $\psi(2\rm S)$ melt near $T_{\rm c}$ (1.1 - 1.2 $T_{\rm c}$)~\cite{satz,satz2}. At LHC energies, due to the large increase of the $c\overline c$ production cross-section with the collision energy, there is a possibility of J/$\psi$ production via recombination of ${c}$ and $\overline {c}$. Thus, the observation of J/$\psi$ production in nucleus-nucleus collisions via recombination also constitutes an evidence of QGP formation. The study of the $\psi(2\rm S)$ production, due to its different binding energy, is complementary to that of the J/$\psi$ and it may also be useful for the evaluation of the temperature of the medium. The $\psi(2\rm S)$-to-J/$\psi$ cross-section ratio is predicted to be very sensitive to the details of the recombination mechanism. Experimentally this ratio is interesting as most of the systematic uncertainties cancel, with the remaining systematic uncertainties being only due to the signal extraction and the efficiency evaluation. The pp results for the charmonium provide a baseline for the nuclear modification factor of charmonium production in \mbox{p-Pb} and \mbox{Pb-Pb} collisions. The study of charmonia in \mbox{p-Pb} collisions can be used as a tool for a quantitative investigation of the cold nuclear matter (CNM) effects including various mechanisms such as gluon shadowing, $c\overline c$ break-up via interaction with nucleons, initial/final state energy loss, relevant in the context of studies of the strong interaction. The region of very small $x$ is accessible at the LHC and therefore strong shadowing and coherent energy loss effects are expected.

\section{ALICE detector and data samples}
The ALICE Collaboration has studied $\psi$(2S) production through its dimuon decay channel, in the Muon Spectrometer which covers the pseudorapidity range $-$4 $< \eta <$ $-$2.5. The ALICE detector is described in detail in~\cite{jinst}. The pp analysis has been performed on a triggered event sample corresponding to an integrated luminosity of $\mathcal{L}^{\rm pp}_{\rm int}$ = 1.35 $\pm$ 0.07 pb$^{-1}$ in the rapidity interval $2.5 < y_{\rm lab} < 4$ at $\sqrt{s}$ = 7 TeV. The p-Pb data have been collected at $\sqrt{s_{\rm NN}}$ = 5.02 TeV under two different configurations, inverting the direction of the p and Pb beams. In this way both forward rapidity $2.03 < y_{\rm cms} < 3.53$ ($\mathcal{L}^{\rm pPb}_{\rm int}$ = 5.01 $\pm$ 0.19 nb$^{-1}$) and backward rapidity $-4.46 < y_{\rm cms} < -2.96$ ($\mathcal{L}^{\rm Pbp}_{\rm int}$ = 5.81 $\pm$ 0.18 nb$^{-1}$) could be accessed, with the positive $y$ defined in the direction of the proton beam. Finally, the Pb-Pb analysis has been performed at $\sqrt{s_{\rm NN}}$ = 2.76 TeV ($\mathcal{L}^{\rm PbPb}_{\rm int}$ = 68.8 $\pm$ 0.9 $\mu$b$^{-1}$) in the rapidity region $2.5 < y_{\rm lab} < 4$.
\section{Results}

\subsection{pp collisions}
\begin{figure}[ht]
\includegraphics[scale=0.38]{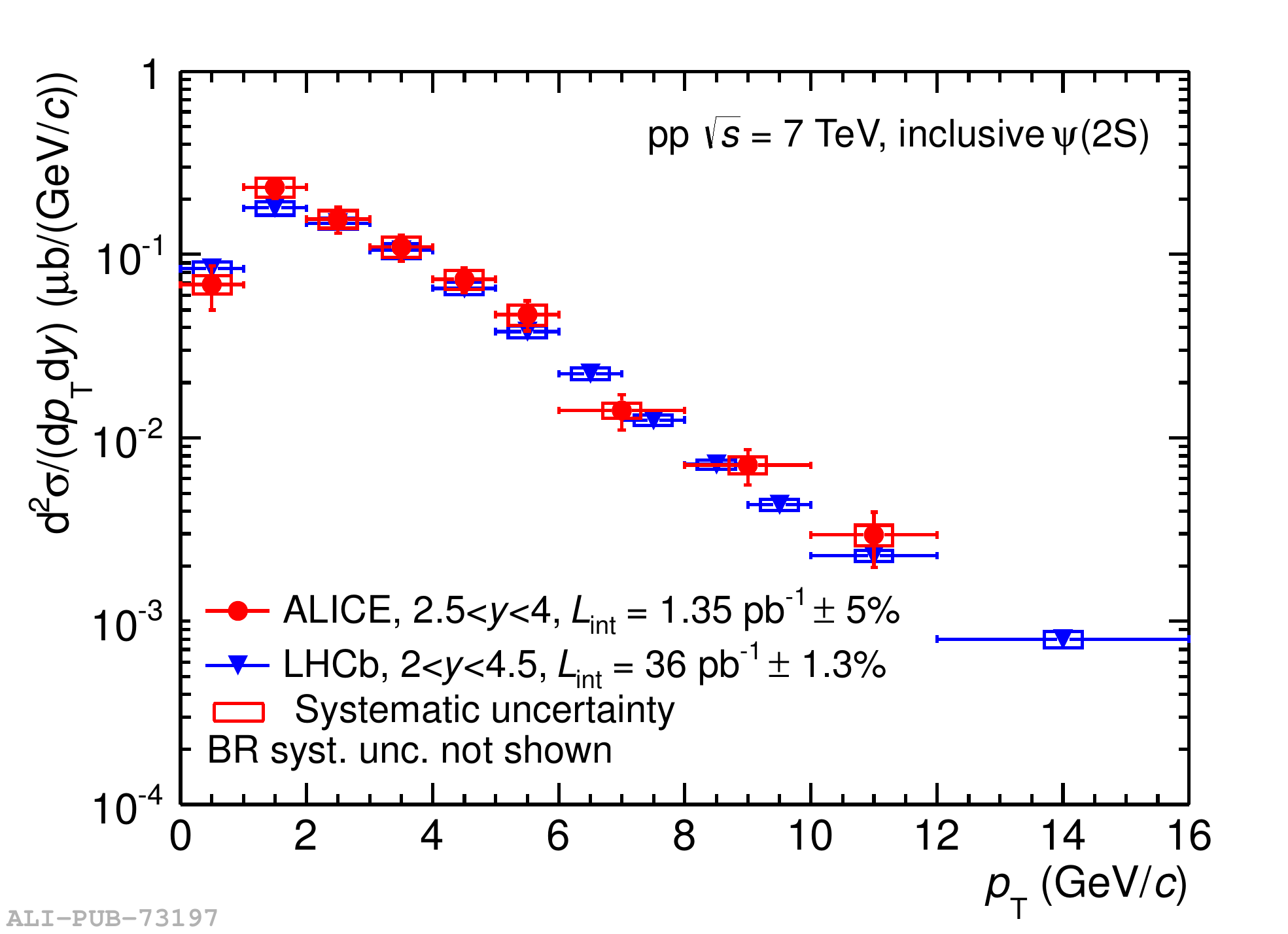}
\includegraphics[scale=0.38]{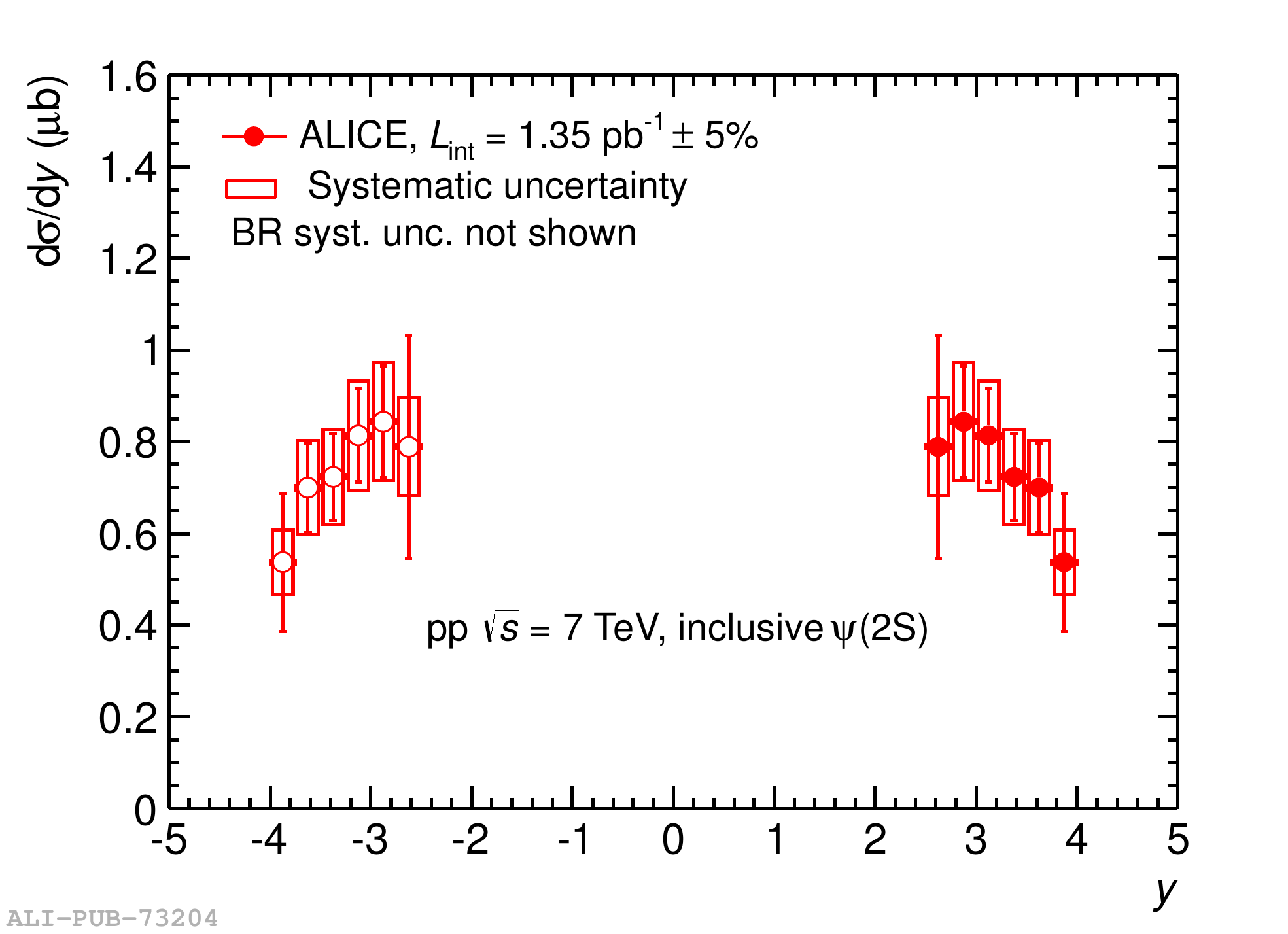}
\caption{\label{fig1}Inclusive differential production cross-sections of $\psi$(2S) as a function of $p_{\rm T}$ (left) and $y$ (right).}
\end{figure}

Fig.~\ref{fig1} shows the inclusive differential production cross-sections of $\psi$(2S) as a function of $p_{\rm T}$ and $y$~\cite{epjc74}. The result on $p_{\rm T}$ differential  cross-section is consistent with the LHCb measurement~\cite{3epjc72} in the same rapidity interval. This is the first measurement of $\psi$(2S) differential cross-sections as a function of $y$ in pp collisions at $\sqrt{s} = 7$ TeV.

\begin{figure}[ht]
\includegraphics[scale=0.38]{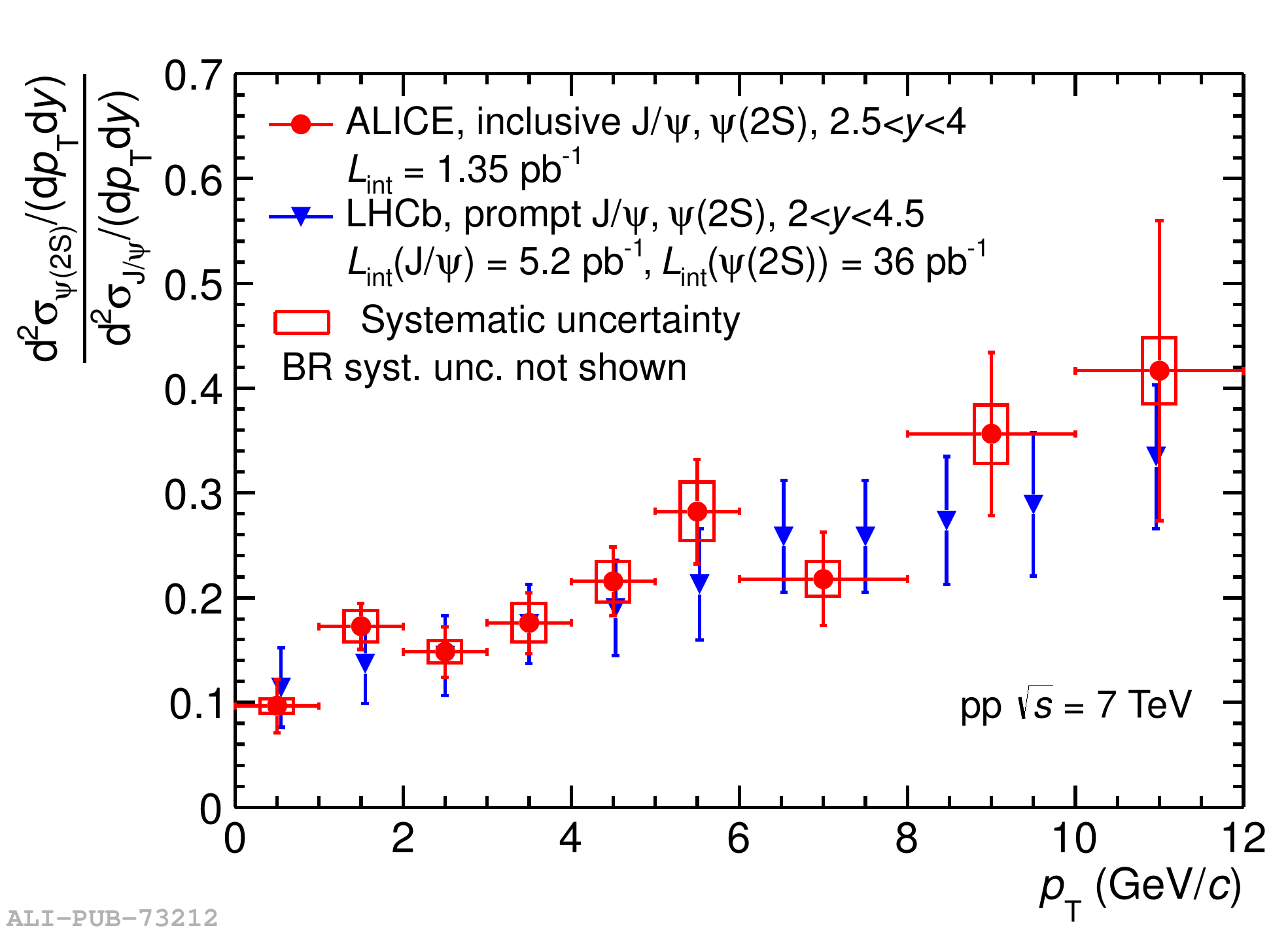}
\includegraphics[scale=0.38]{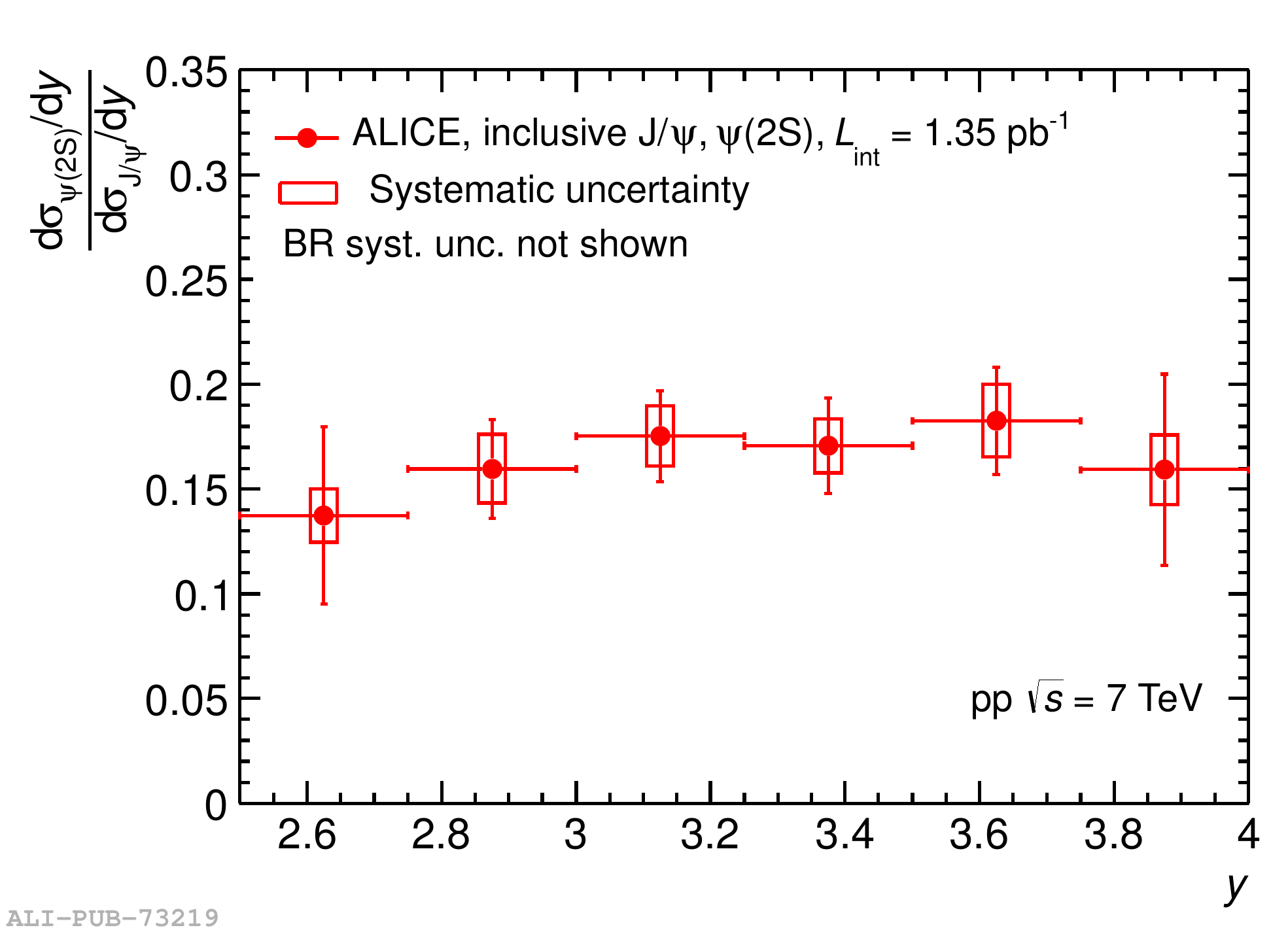}
\caption{\label{fig2}Inclusive $\psi$(2S) to J/$\psi$ cross-section ratio as a function of $p_{\rm T}$ (left) and $y$ (right).}
\end{figure}

The inclusive $\psi$(2S) to J/$\psi$ cross-section ratio was measured as a function of $p_{\rm T}$ and $y$ as shown in Fig.~\ref{fig2}. A clear $p_{\rm T}$  dependence can be observed, consistent with the one measured by LHCb~\cite{3epjc72}. No strong $y$ dependence is visible in the $y$ range covered by the ALICE muon spectrometer.  

\subsection{p-Pb collisions}
\begin{figure}[ht]
\includegraphics[scale=0.38]{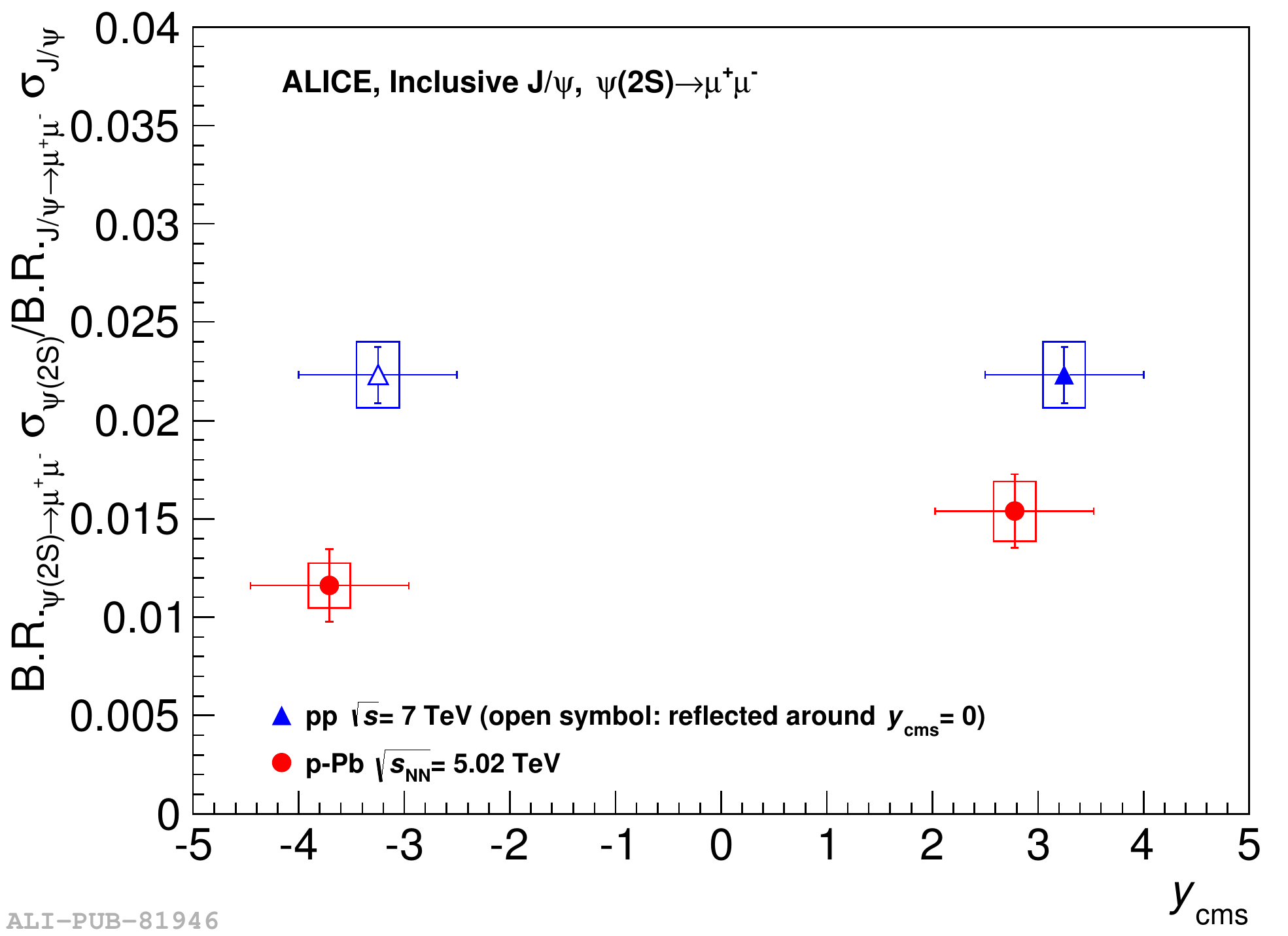}
\includegraphics[scale=0.38]{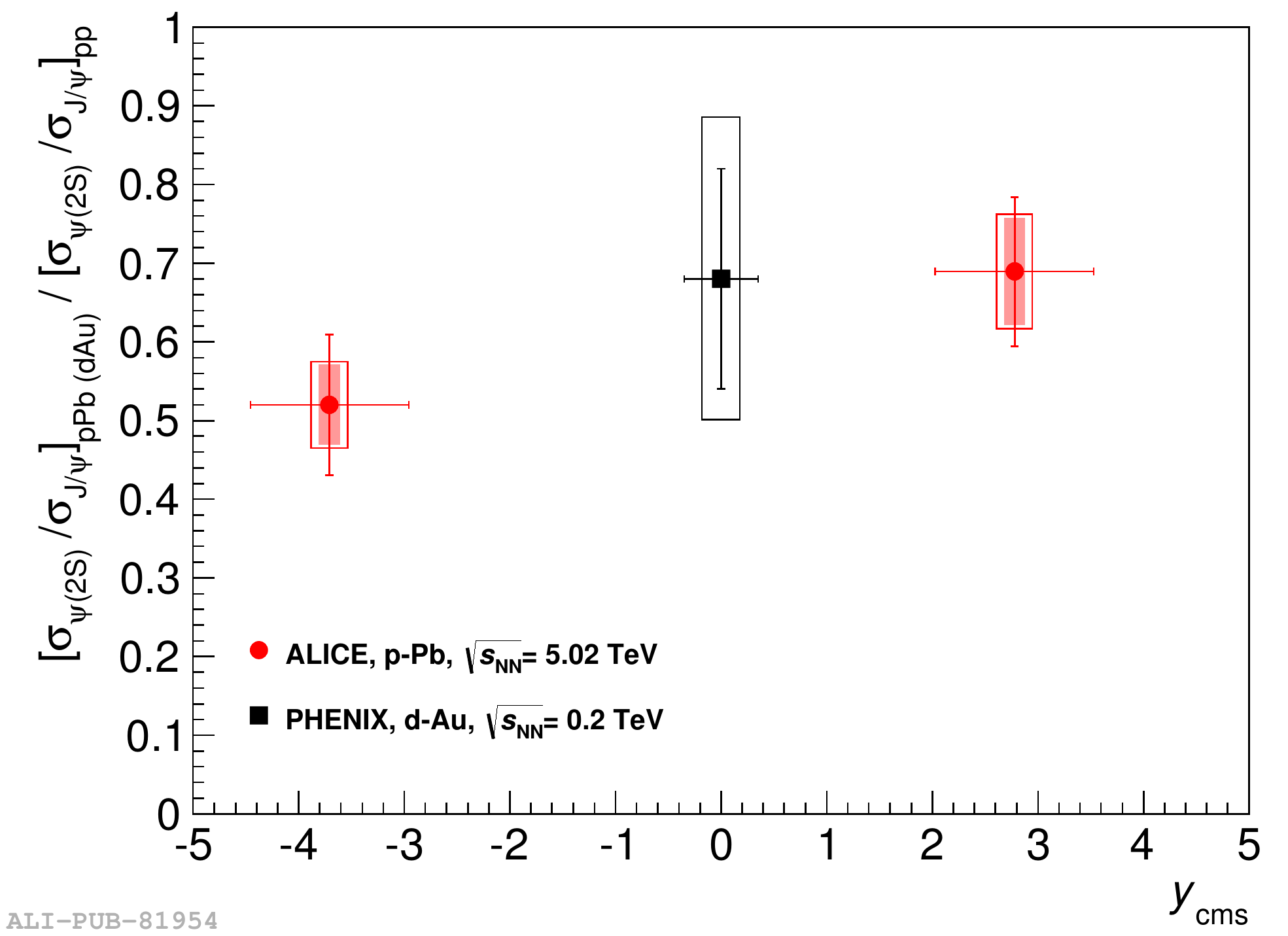}
\caption{\label{fig3}Left: $\psi$(2S) to J/$\psi$ cross-section ratio compared to the pp results at $\sqrt{s}$ = 7 TeV. Right: the double ratio compared to the PHENIX result~\cite{ada13}.}
\end{figure}

The production cross section of $\psi$(2S) in p-Pb is compared to the J/$\psi$ one and to the corresponding quantities in pp collisions at $\sqrt{s}$ = 7 TeV (no LHC results are available at $\sqrt{s}$ = 5.02 TeV) using the $\psi$(2S) to J/$\psi$ ratio and the double ratio [$\sigma_{\rm \psi(2S)}/\sigma_{\rm J/\psi}]_{\rm pPb}$/[$\sigma_{\rm \psi(2S)}/\sigma_{\rm J/\psi}]_{\rm pp}$~\cite{jhep12,arnaldi} as shown in Fig.~\ref{fig3}. The pp ratios are significantly higher than those for p-Pb and Pb-p. The double ratio is compared with the corresponding measurement by the PHENIX Collaboration at mid-rapidity at $\sqrt{s_{\rm NN}}$ = 0.2 TeV~\cite{ada13}. Within uncertainties, a similar relative $\psi$(2S) suppression is observed by the two experiments.

\begin{figure}[ht]
\centering
\includegraphics[scale=0.38]{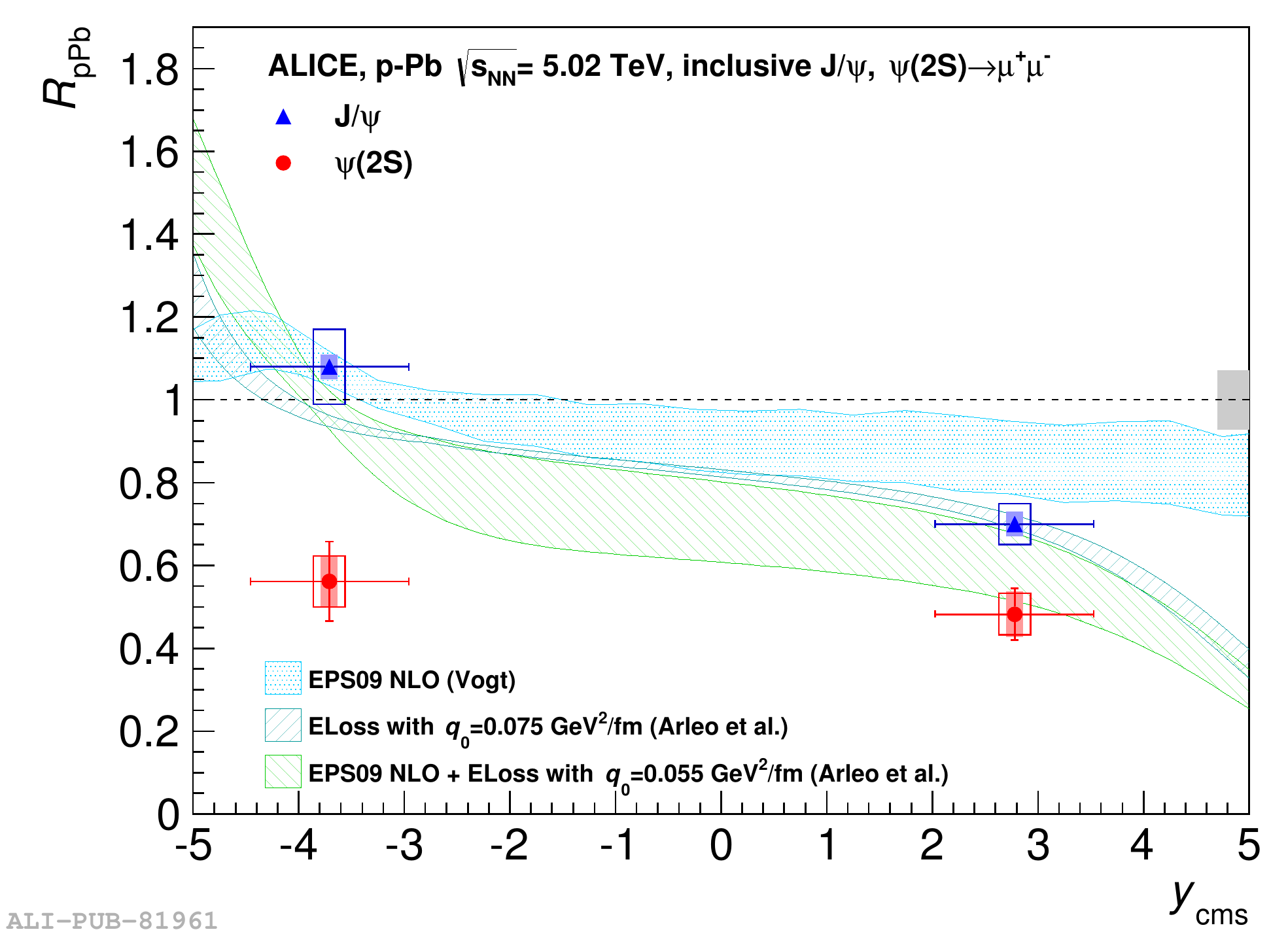}
\caption{\label{fig4}$\psi$(2S) $R_{\rm pPb}$ versus $y$ compared to the J/$\psi$ $R_{\rm pPb}$ and theoretical models.}
\end{figure}

Since no result on cross-section of $\psi$(2S) is available at $\sqrt{s}$ = 5.02 TeV in pp collisons, the nuclear modification factor of $\psi$(2S) is obtained by combining the J/$\psi$ $R_{\rm pPb}$~\cite{jhep1402} and the double ratio, as $R_{\rm pPb}^{\rm \psi(2S)} =$$ R_{\rm pPb}^{\rm J/\psi}$$\times$$(\sigma_{\rm pPb}^{\rm \psi(2S)}/\sigma_{\rm pPb}^{\rm J/\psi})$$\times$$(\rm \sigma_{\rm pp}^{\rm J/\psi}/\sigma_{\rm pp}^{\rm \psi(2S)})$, assuming that the ratio in pp collisions does not depend on $\sqrt{s}$~\cite{jhep12}. In Fig.~\ref{fig4}, $R_{\rm pPb}^{\psi(2S)}$ is compared with $R_{\rm pPb}^{J/\psi}$ and also with theoretical calculations based on nuclear shadowing~\cite{ijmp}, coherent energy loss or both~\cite{jhep1303}. The suppression of $\psi$(2S) production is stronger than that of J/$\psi$ and reaches a factor of 2 with respect to pp. Since the kinematic distributions of gluons producing the J/$\psi$ or the $\psi$(2S) are rather similar and since the coherent energy loss does not depend on the final quantum numbers of the resonances, the same theoretical calculations hold for both J/$\psi$ and $\psi$(2S). Theoretical models predict a $y$ dependence which are in reasonable agreement with the J/$\psi$ results but no model can describe the $\psi$(2S) data. These results show that other mechanisms must be invoked in order to describe the $\psi$(2S) suppression in p-Pb collisions.

\begin{figure}[ht]
\includegraphics[scale=0.38]{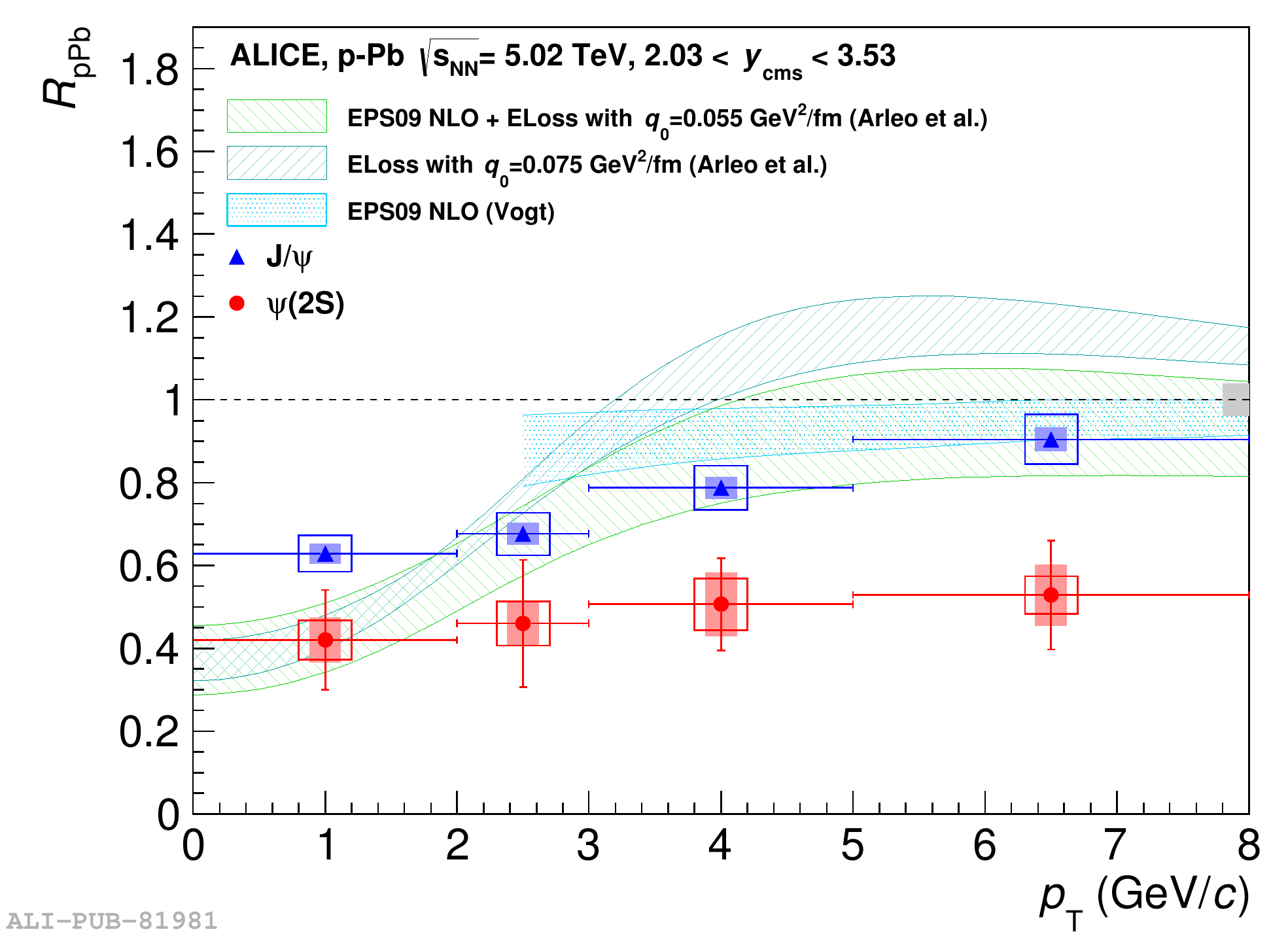}
\includegraphics[scale=0.38]{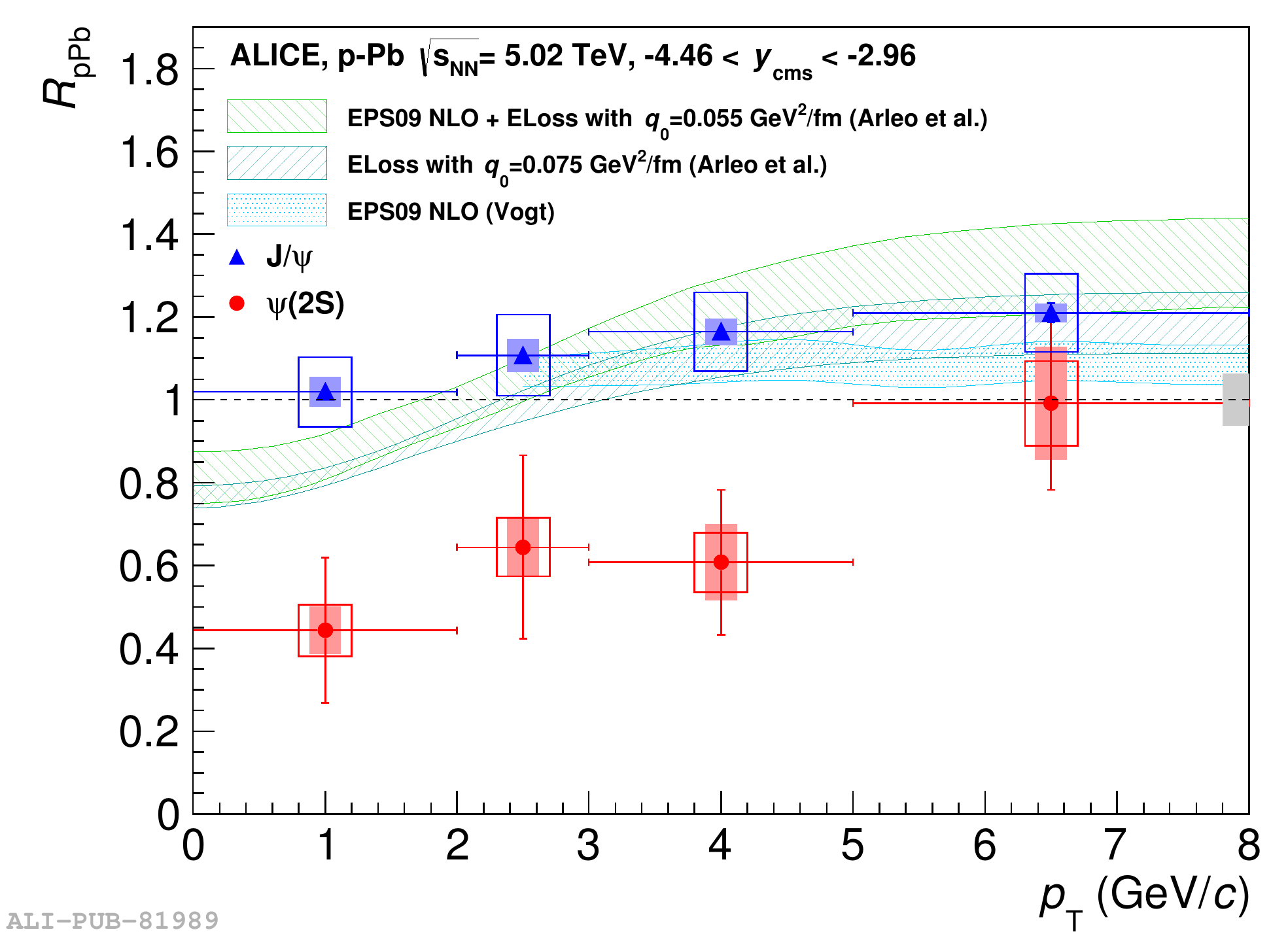}
\caption{\label{fig5}$p_{\rm T}$ dependence of the $\psi$(2S) $R_{\rm pPb}$ compared to the J/$\psi$ $R_{\rm pPb}$ and theoretical calculations in the forward (left) and backward (right) rapidity region.}
\end{figure}

The $R_{\rm pPb}$ is also computed as a function of $p_{\rm T}$ both at backward and forward $y$ and the results are shown in Fig.~\ref{fig5}. At both rapidities, the $R_{\rm pPb}^{\rm \psi(2S)}$ shows a strong suppression with a slightly more evident $p_{\rm T}$ dependence at backward-$y$. The $\psi$(2S) is more suppressed than the J/$\psi$, as already observed for the $p_{\rm T}$-integrated result. Theoretical calculations are in fair agreement with the $R_{\rm pPb}^{\rm J/\psi}$ but clearly overestimate the $R_{\rm pPb}^{\rm \psi(2S)}$ behaviour. The calculations from the comover model~\cite{ferre} shows that the interaction with comovers, mostly at play in the backward region, is able to explain the stronger $\psi$(2S) suppression. 

\begin{figure}[ht]
\includegraphics[scale=0.38]{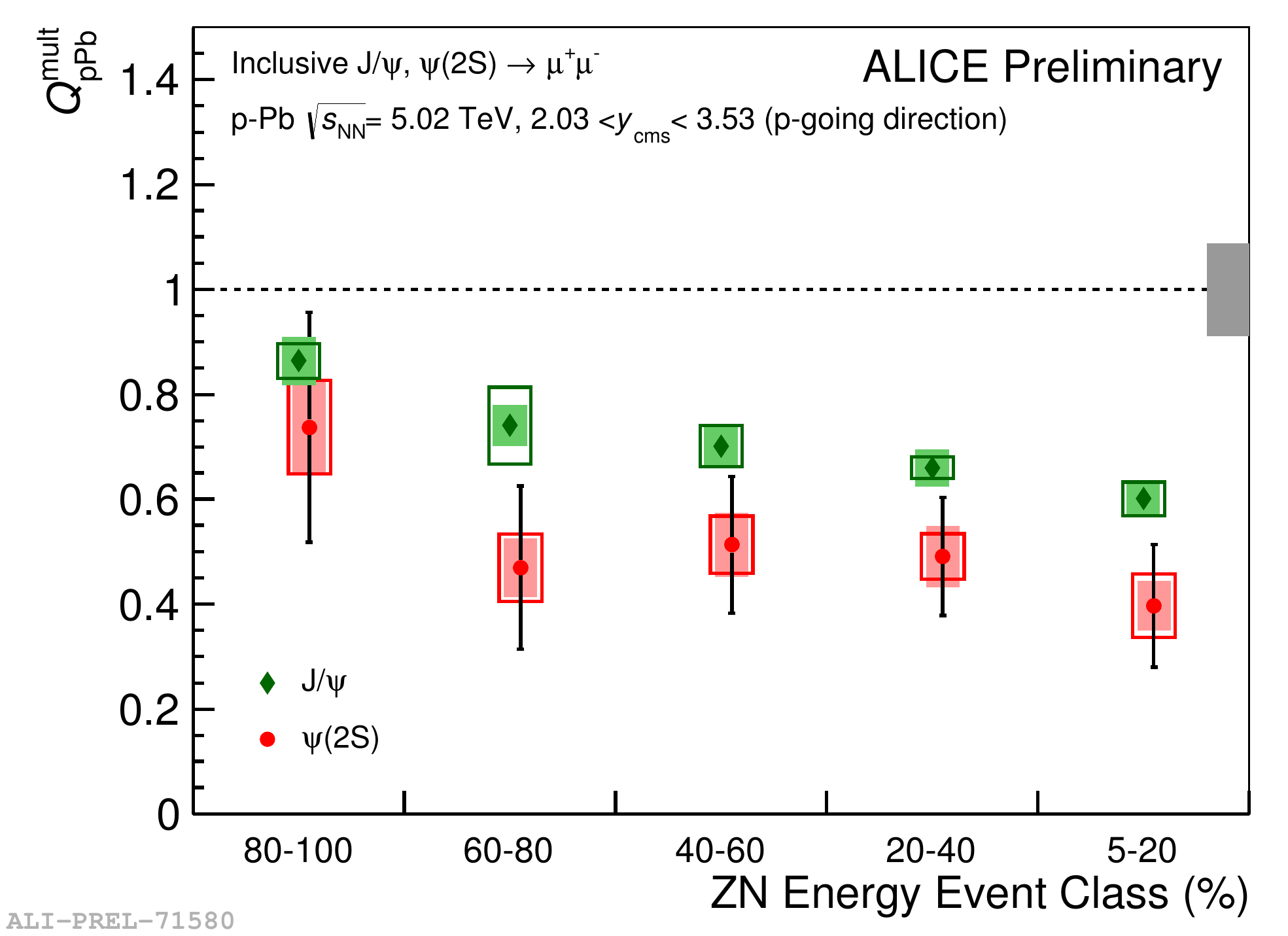}
\includegraphics[scale=0.38]{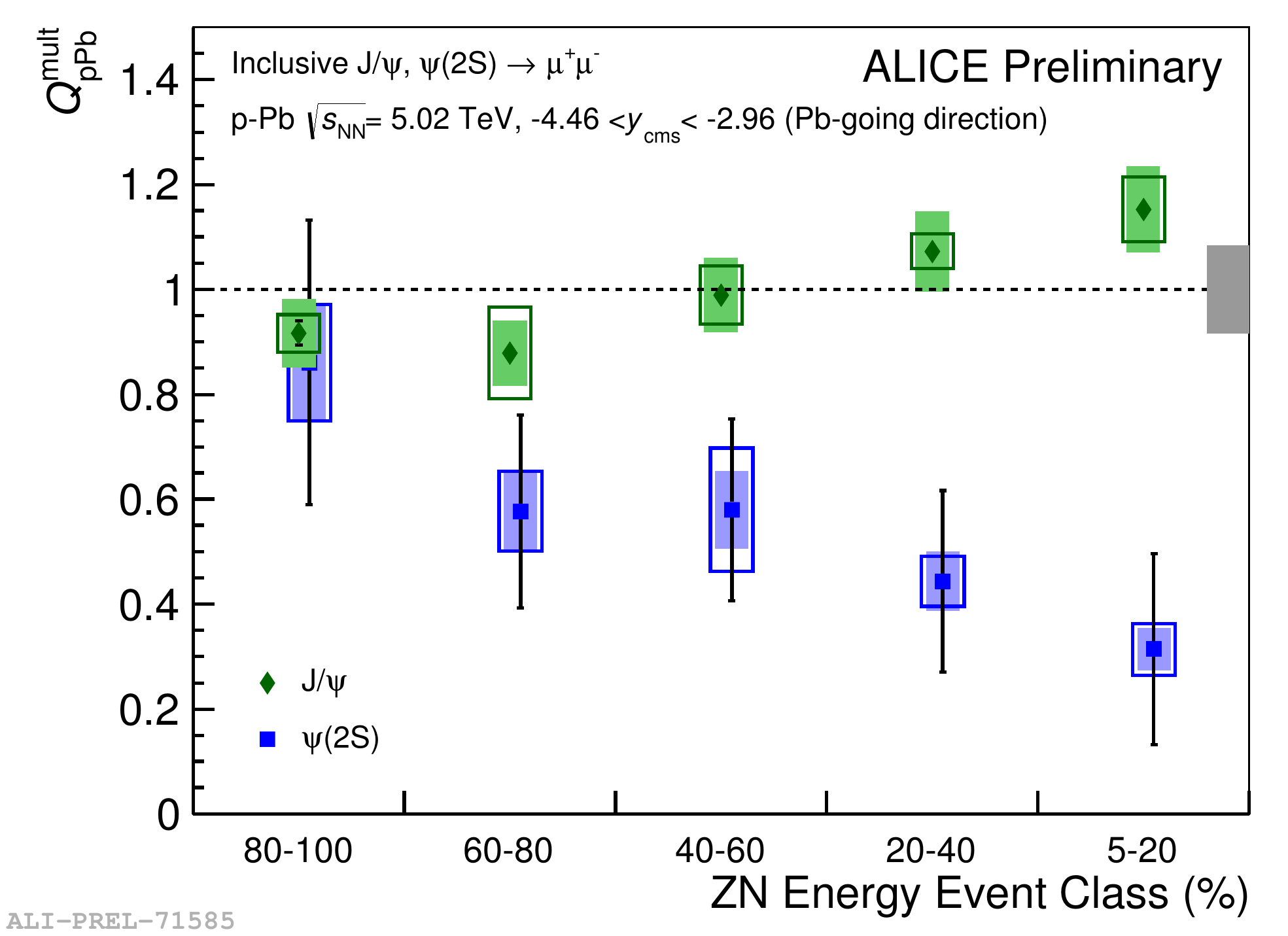}
\caption{\label{fig6}$\psi$(2S) $Q_{\rm pPb}$ versus event activity compared to the J/$\psi$ $Q_{\rm pPb}$ in the forward (left) and backward (right) rapidity region.}
\end{figure}

Finally, the $\psi$(2S) production is studied as a function of the collision event activity both at backward and forward $y$~\cite{arnaldi}, as shown in Fig.~\ref{fig6}. The event activity determination is described in details in~\cite{toia}. Since the centrality determination in p-Pb collisions can be biased by the choice of the estimator, the nuclear modification factor is, in this case, named $Q_{\rm pPb}$~\cite{toia}. The $\psi$(2S) $Q_{\rm pPb}$ shows a strong suppression, which increases with increasing event activity, and is rather similar in both the forward and the backward $y$ regions. The J/$\psi$ $Q_{\rm pPb}$ shows a similar decreasing trend at forward-$y$ as a function of the event activity. On the contrary, the J/$\psi$ and $\psi$(2S) $Q_{\rm pPb}$ patterns observed at backward-$y$ are rather different, with the $\psi$(2S) significantly more suppressed for large event activity classes.

\subsection{Pb-Pb collisions}
\begin{figure}[ht]
\centering
\includegraphics[scale=0.40]{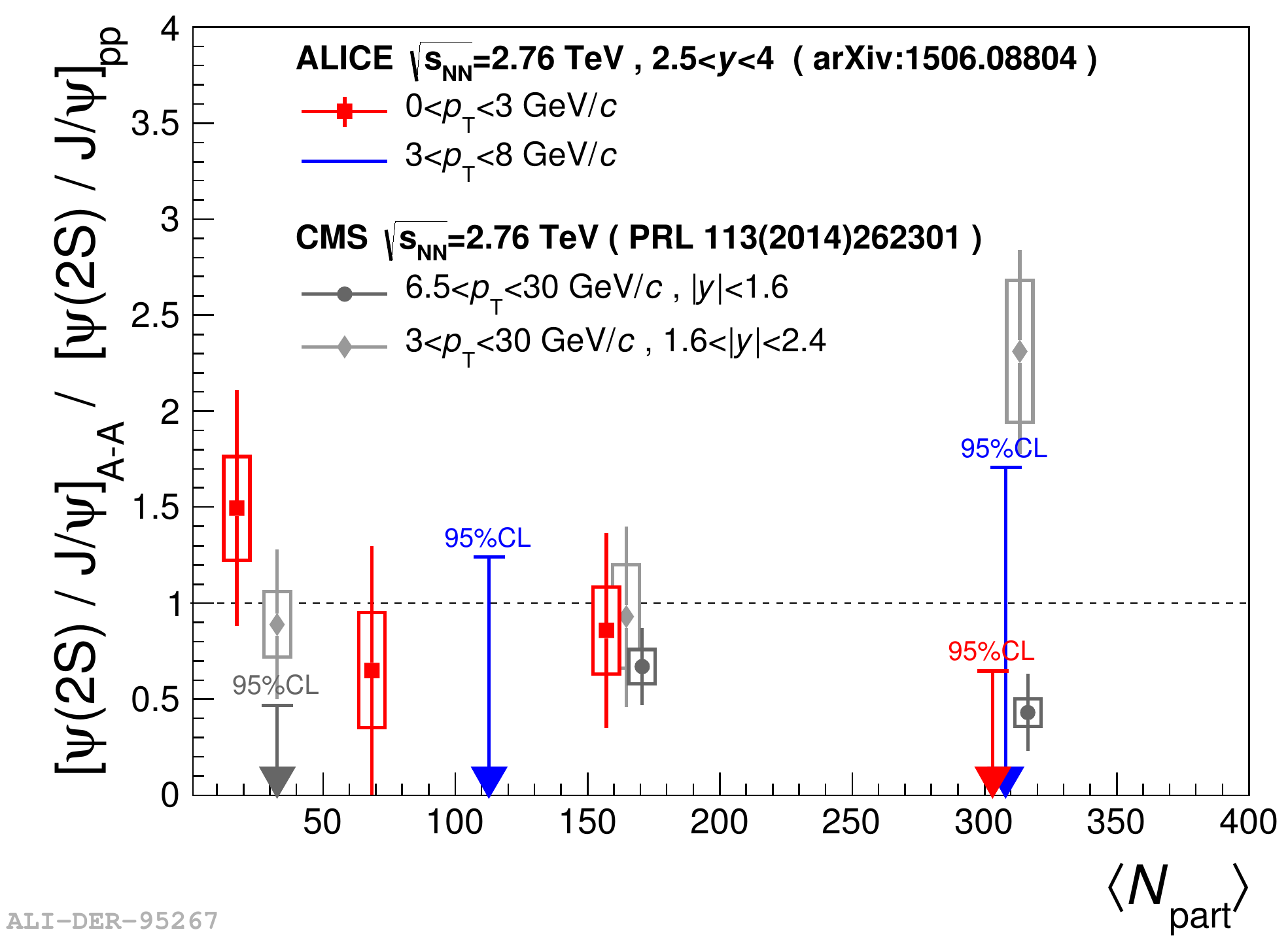}
\caption{\label{fig7}Double ratio [${\rm \psi(2S)}/{\rm J/\psi}]_{\rm PbPb}$/[${\rm \psi(2S)}/{\rm J/\psi}]_{\rm pp}$ as a function of centrality in two $p_{\rm T}$ intervals compared to CMS measurements~\cite{prl113}.}
\end{figure}

In order to study the suppression of $\psi$(2S) relative to J/$\psi$ in Pb-Pb collisions, the double ratio [$\sigma_{\rm \psi(2S)}/\sigma_{\rm J/\psi}]_{\rm PbPb}$/[$\sigma_{\rm \psi(2S)}/\sigma_{\rm J/\psi}]_{\rm pp}$ has been measured as a function of centrality in two $p_{\rm T}$ intervals (0 $<$ $p_{\rm T}$ $<$ 3 GeV/$\it{c}$ and 3 $<$ $p_{\rm T}$ $<$ 8 GeV/$\it{c}$)~\cite{arnaldi13} and has been compared with the results from the CMS Collaboration~\cite{prl113} as shown in Fig.~\ref{fig7}. Limited $\psi$(2S) statistics does not allow any firm conclusion about the centrality dependence of this ratio and the comparison with CMS is not straightforward due to the different kinematic coverage.

\section{Summary}

In summary, the ALICE Collaboration has studied the inclusive $\psi$(2S) production in pp, p-Pb and Pb-Pb collisions at the LHC. The $\psi$(2S) production cross-section and the $\psi$(2S) to J/$\psi$ cross-section ratio have been obtained as a function of $p_{\rm T}$ and $y$ in pp collisions. The $p_{\rm T}$ differential results are in good agreement with LHCb measurements. In p-Pb collisions the results show that $\psi$(2S) is significantly more suppressed than J/$\psi$ in both rapidity regions. This observation implies that initial state nuclear effects alone cannot account for the modification of the $\psi$(2S) yields, as also confirmed by the poor agreement of the nuclear modification factor of $\psi$(2S) with models based on shadowing and/or energy loss. Interaction with comovers is able to explain the $\psi$(2S) suppression. The final state interaction with the hadronic medium could also provide a possible explanation for the stronger $\psi$(2S) suppression~\cite{rapp}. Limited statistics prevent to make definitive conclusions on $\psi$(2S) production in Pb-Pb collisions.

\end{document}